# Magnon-driven dynamics of frustrated skyrmion in synthetic antiferromagnets: Effect of skyrmion precession

Z. Jin[1], Y. F. Hu[1], T. T. Liu[1], Y. Liu[1], Z. P. Hou[1], D. Y. Chen[1], Z. Fan[1], M. Zeng[1], X. B. Lu[1], X. S. Gao[1], M. H. Qin[1*], and J. –M. Liu[1,2]

[1]*Guangdong Provincial Key Laboratory of Quantum Engineering and Quantum Materials and Institute for Advanced Materials, South China Academy of Advanced Optoelectronics, South China Normal University, Guangzhou 510006, China*

[2]*Laboratory of Solid State Microstructures, Nanjing University, Nanjing 210093, China*

**[Abstract]** A theoretical study on the interplay of frustrated skyrmion and magnons is useful for revealing new physics and future experiments design. In this work, we investigated the magnon-driven dynamics of frustrated skyrmion in synthetic antiferromagnets, focusing on the effect of skyrmion precession. It is theoretically revealed that the scattering cross section of the injected magnons depends on the skyrmion precession, which in turn effectively modulates the skyrmion Hall motion. Specifically, the Hall angle decreases as the precession speed increases, which is also verified by the atomistic micromagnetic simulations. Moreover, the precession speed and the Hall angle of the frustrated skyrmion depending on the magnon intensity and damping constant are simulated, demonstrating the effective suppression of the Hall motion by the skyrmion precession. This work provides a comprehensive understanding of the magnon-skyrmion scattering in frustrated magnets, benefiting future spintronic and magnonic applications.

Email: qinmh@scnu.edu.cn

# I. Introduction

Magnetic skyrmions [1] which are topologically vortex-like spin configurations have attracted extensive attention since their first discovery in B20 compound [2], due to their interesting physics and potential applications in skyrmion-based spintronic devices such as race-track memory and logic units. Ferromagnetic skyrmions have been observed in a series of chiral magnets [2-6] and heavy metal/ferromagnetic films [7-10] with inversion broken symmetry, attributing to the indispensable Dzyaloshinskii-Moriya interactions (DMI). Moreover, it has been proven that the skyrmions can be effectively driven by various stimuli including spin-polarized electric current [11,12], gradient magnetic field [13], gradient electric and oscillating field [14,15].

Most recently, it was predicted theoretically [13,16-20] and discovered experimentally [21,22] that the skyrmions can be stabilized by the competing interactions in frustrated magnets with inversion symmetry in the absence of the DMI. Compared to the skyrmions stabilized by the DMI, those in frustrated magnets have a new degrees-of-freedom associated with the rotation of helicity due to the fact that the skyrmion energy is independent on the direction of spin rotation [17]. Interestingly, the rotation of helicity couples to the usual translational motion of the skyrmion driven by the spin Hall effect induced by the out-of-plane current, resulting in the rotational skyrmion motion [19,23]. It is suggested that the rotational motion of skyrmions can be harnessed for building nano-oscillators.

Compared to the electric current, magnons as the quanta of spin waves, driving skyrmion motion without Joule heating due to the absence of charge physical transport, are particularly attractive for the advantage of low-energy consumption [24-33]. For example, the magnons in ferromagnets are only right-circularly polarized and deflected by the effective field from the skyrmions, which in turn drives the skyrmion motion towards the spin wave source accompanied by Hall motion through the momenta exchange [28,31,34]. Moreover, circularly polarized magnons are also deflected by antiferromagnetic (AFM) skyrmions toward transverse direction, and consequently drive the skyrmion Hall motion even in antiferromagnets [32,35].

These important works unveil interesting electric driven skyrmion dynamics in frustrated magnets and magnon driven skyrmion dynamics in ferromagnets and antiferromagnets, while the dynamic properties of frustrated skyrmions driven by polarized magnons remain elusive. On one hand, it is expected that the translational motion of skyrmion probably couples to the precession mode as magnons pass through frustrated skyrmion. Moreover, the rotation of helicity may affect the skyrmion-magnon scattering potential, and effectively modulates the skyrmion dynamics. As a matter of fact, a strong dependence of magnon-domain wall scattering on the precession speed of the AFM domain wall has been revealed recently [36]. With the increase of the precession speed, the scattering potential well is replaced by the potential barrier, resulting in the reflection of the magnons and the forward motion of the wall. In some extent, similar phenomenon could be available in the magnon driven skyrmion dynamics in frustrated magnets, considering the highly relevance between these magnetic structures.

On the other hand, for a precessing AFM skyrmion which is a Newtonian particle rotating around an axis, the angular momentum is transferred from the injected magnons to the skyrmion, definitely affecting the whole scattering process. Thus, the clarification of the frustrated skyrmion dynamics driven by polarized magnons can help one to select potential materials for future experiments and device design, besides its benefit to the development of spintronics and newly rising magnonics.

In this work, we study the skyrmion dynamics in antiferromagnetically coupled frustrated bilayers driven by circularly polarized magnons. It is analytically revealed that the scattering cross section of the magnons depends on the skyrmion precession speed, which in turn determines the skyrmion Hall angle. The prediction is verified by the numerical simulations of the atomistic spin model based on the Landau-Lifshitz-Gilbert (LLG) equation. Moreover, the effective modulation of the Hall angle by the magnon intensity and damping constant is also revealed, providing useful information for future spintronic and magnonic applications based on frustrated magnets.

## II. Model and methods

Following the earlier work, we consider two frustrated $FM^1/FM^2$ layers with competing $J_1$-$J_2$-$J_3$ Heisenberg interactions and the magnetic moments $\mathbf{m}^1/\mathbf{m}^2$ satisfying condition $|\mathbf{m}^1| = |\mathbf{m}^2|$

= S with the spin length S. The two layers are coupled by an AFM interfacial exchange coupling $J_{\text{inter}}$ between the nearest neighbors which can be realized by utilizing the Ruderman-Kittel-Kasuya-Yosida interaction. Similarly, the total magnetization $\mathbf{m} = (\mathbf{m}^1 + \mathbf{m}^2)/2S$ and normalized staggered Néel vector $\mathbf{n} = (\mathbf{m}^1 - \mathbf{m}^2)/2S$ are defined to describe the skyrmion dynamics. For a fixed or slowly varying spin texture, the continuum Lagrangian density of frustrated magnets reads (modeling details, parameters and the detailed derivation are presented in Appendix A)

$$L = \frac{\rho^2}{2A_0}\dot{\mathbf{n}}^2 - \frac{A}{2}(\nabla \mathbf{n})^2 + \frac{K}{2}n_z^2, \tag{1}$$

where $A_0$ and $A$ are the homogeneous and effective exchange constants respectively [37,38], $K$ is the easy $z$-axis anisotropy constant, and $\rho = \hbar S/a$ is the density of the staggered spin angular momentum per unit cell [39] with the lattice constant $a$ and reduced Planck's constant $\hbar$.

To describe the magnons, we use a global frame defined by three mutually orthogonal unit vectors ($\mathbf{e}_1$, $\mathbf{e}_2$, $\mathbf{e}_3$) with $\mathbf{e}_3 = \mathbf{n}_0/|\mathbf{n}_0| = \mathbf{e}_1 \times \mathbf{e}_2$ with the equilibrium configuration $\mathbf{n}_0$ [40,41]. Thus, an excited state can be parametrized as $\mathbf{n} = \mathbf{n}_0 + \delta_1\mathbf{e}_1 + \delta_2\mathbf{e}_2$ with $\delta_1$ and $\delta_2$ describing the amplitude components of the magnon. Then, two monochromatic solutions with the complex fields $\psi = (\delta_1 \pm i\delta_2)/\sqrt{2}$ for right/left circularly polarized magnon modes are obtained, corresponding to the anti-clockwise/clockwise precession of $\mathbf{n}$. Subsequently, the magnon-skyrmion scattering are analytically calculated using the elastic scattering theory, which has been successfully applied in studying the skyrmion dynamics in chiral magnets [30].

Moreover, the skyrmion Hall angle as a function of the precession velocity is also numerically estimated using the LLG simulations of the discrete spin model, in order to check the validity of the theoretical analysis. The simulation details are presented in Appendix B.

### III. Results and discussion

*A. Magnons scattered by skyrmion without precession*

First, we study the skyrmion-magnon scattering in the absence of skyrmion precession. In this case, the skyrmion works as a fictitious electromagnetic field and induces the magnon deviation from its propagation direction. The spin wave fluctuations are considered to obtain the scattering potential from the frustrated skyrmion. For the parameterization of the magnons, the local orthogonal frame ($\mathbf{e}_1$, $\mathbf{e}_2$, $\mathbf{e}_3$) is represented by the polar angle $\theta$ and azimuthal angle $\varphi$ of the local spin, with $\mathbf{e}_1 = (-\sin\varphi, \cos\varphi, 0)$, $\mathbf{e}_2 = (-\cos\theta\cos\varphi, -\cos\theta\sin\varphi, \sin\theta)$ and $\mathbf{e}_3 = (\sin\theta\cos\varphi, \sin\theta\sin\varphi, \cos\theta)$, where $\theta = \theta(r)$ is related to the position vector $\mathbf{r}$ with the length $r$ and azimuthal angle $\chi$.

The remaining massive fluctuation modes are represented by the dimensionless complex field $\psi$ and corresponding spinor notation $\vec{\psi} = (\psi, \psi^*)^T$, and the Néel vector is expressed as [30]

$$n_x = \sqrt{1-2|\psi|^2}\sin\theta\cos\varphi + \frac{1}{\sqrt{2}}(-\sin\varphi - i\cos\theta\cos\varphi)\psi + \frac{1}{\sqrt{2}}(-\sin\varphi + i\cos\theta\cos\varphi)\psi^*$$
$$n_y = \sqrt{1-2|\psi|^2}\sin\theta\sin\varphi + \frac{1}{\sqrt{2}}(\cos\varphi - i\cos\theta\sin\varphi)\psi + \frac{1}{\sqrt{2}}(\cos\varphi + i\cos\theta\sin\varphi)\psi^* \quad , \quad (2)$$
$$n_z = \sqrt{1-2|\psi|^2}\cos\theta + \frac{i}{\sqrt{2}}\sin\theta\psi - \frac{i}{\sqrt{2}}\sin\theta\psi^*$$

For weak fluctuations, we expand the Lagrangian in the vicinity of the skyrmion solution to the second order and invoke the Euler-Lagrangian rule. Then, the magnon dynamics is obtained as the eigenstates of the Klein-Gordon equation $H\vec{\psi} = -\rho^2 \mathbf{I} \partial_t^2 \vec{\psi} / A_0$ with $H = H_0 + H_s$ and the identity matrix $\mathbf{I}$. Here, the ground-state Hamiltonian $H_0$ describes the magnons in the absence of the skyrmion

$$H_0 = A[\mathbf{I}(-\partial_r^2 - \frac{\partial_r}{r} + \frac{-\partial_\chi^2 + 1}{r^2} + \frac{K}{A}) + \tau^z \frac{2i\partial_\chi}{r^2}], \qquad (3)$$

and the skyrmion matrix scattering potential $H_s$ is given by

$$H_s = A(U_z \tau^z + U_0 \mathbf{I} + U_x \tau^x), \qquad (4)$$

where

$$U_z = 2i\partial_\chi(\frac{\cos\theta - 1}{r^2})$$
$$U_0 = \frac{3(\cos 2\theta - 1)}{4r^2} - \frac{\partial_r\theta^2}{2} - \frac{K(\sin^2\theta - 2\cos^2\theta + 2)}{2A}, \qquad (5)$$
$$U_x = \frac{\sin^2\theta}{2r^2} - \frac{\partial_r\theta^2}{2} + K\frac{\sin^2\theta}{2A}$$

with the Pauli matrix $\tau^z$ and $\tau^x$.

Thus, a skew scattering of the magnons is expected due to the scattering potentials from the skyrmion. For high energy, as an example, the potential $U_z$ works as a fictitious magnetic flux and deflects the magnons, resulting in the magnon Hall effect as the magnons pass through the skyrmion. It is worth noting that similar scattering potentials have been derived in ferromagnets, where the skew scattering of the circularly polarized magnons by the ferromagnetic skyrmion is revealed [30]. These analogous results partially attribute to the similarity in Hamiltonians between ferromagnets and synthetic antiferromagnets, where the magnetization and the Néel vector play respectively a similarly role [42]. However, in the latter system, there exist two degenerate magnon modes with circular polarizations in the absence of magnetic fields, i.e. left- ($\omega > 0$) and right- ($\omega < 0$) circularly polarized modes, due to the additional time-reversal symmetry combined with exchanging the two sublattices. As a result, the eigen-equation in the synthetic antiferromagnets is the Klein-Gordon equation rather than Schrödinger-like equation.

In addition, the precession of the frustrated skyrmions has been reported. Similar to the spin-orbit coupling of light [43], the skyrmion precession mode may couple to the spin angular momentum of the magnons and in turn affects the entire scattering process, as will be revealed in the next section.

*B. Effect of skyrmion precession*

Considering a uniform precession of the skyrmion, the skyrmion profile can be described by $\varphi = \varphi_0 + \Omega t$ with the angular velocity $\mathbf{\Omega} = \Omega \mathbf{z}$. In this case, $\dot{\mathbf{n}}$ in the rotating frame is updated to $\dot{\mathbf{n}} + \mathbf{\Omega} \times \mathbf{n}$ in the lab frame [36], and the Lagrangian density Eq. (1) changes to

$$L = \frac{\rho^2}{2A_0}(\dot{\mathbf{n}})^2 + \frac{\rho^2}{A_0}\dot{\mathbf{n}}(\mathbf{\Omega}\times\mathbf{n}) + \frac{\rho^2}{2A_0}\Omega^2(\mathbf{z}\times\mathbf{n})^2 + \frac{A}{2}(\nabla n)^2 - \frac{K}{2}(n_z)^2 . \tag{6}$$

Similarly, we obtain the ground state Hamiltonian $H_0$ and the skyrmion potential $H_s$:

$$H_0 = A\{\mathbf{I}[-\partial_r^2 - \frac{\partial_r}{r} + \frac{-\partial_\chi^2 + 1}{r^2} + \frac{(K+\rho^2\Omega^2/A_0)}{A}] + \tau^z(\frac{2\rho^2}{AA_0}\Omega i\partial_t + \frac{2i\partial_\chi}{r^2})\} , \tag{7}$$

$$H_s = A(U_z\tau^z + U_0\mathbf{I} + U_x\tau^x) , \tag{8}$$

with

$$U_z = 2i\partial_\chi(\frac{\cos\theta-1}{r^2}) + 2\frac{\rho^2}{AA_0}\Omega i\partial_t(\cos\theta-1)$$

$$U_0 = \frac{3(\cos 2\theta-1)}{4r^2} - \frac{\partial_r\theta^2}{2} - \frac{(K+\rho^2\Omega^2/A_0)(\sin^2\theta-2\cos^2\theta+2)}{2A}. \quad (9)$$

$$U_x = \frac{\sin^2\theta}{2r^2} - \frac{\partial_r\theta^2}{2} + (K+\rho^2\Omega^2/A_0)\frac{\sin^2\theta}{2A}$$

From Eq. (9), it is noted that the precession enhances the easy-axis anisotropy to $K+\rho^2\Omega^2/A_0$ and induces an additional effective Néel field $2i\Omega\rho^2\partial_t/A_0$ (the field is equivalent to the magnetic field in ferromagnetic system). Thus, the enhanced anisotropy and Néel field set an upper limit to the precession frequency, above which the skyrmion cannot be stabilized anymore. For high energy $\omega^2 \gg K$, $U_z$ term dominates in the scattering potential and is only considered in solving the eigen-equation [30]. Moreover, the effective Néel field is rather larger than the effective anisotropy due to $\Omega^2 \ll 2\omega\Omega$, which could be safely ignored.

Subsequently, we introduce the angular momentum $m$ with $\psi = \psi_m(r)\exp[i(m\chi-\omega t)]$ to classify the eigenmodes [30,44] and the eigen-equation. Then, the eigen-equation, $H_0$ and $U_z$ are updated to

$$(U_z + H_0)\psi_m(r) = \rho^2\omega^2\psi_m(r)/A_0. \quad (10)$$

$$H_0 = A\{-\partial_r^2 - \frac{\partial_r}{r} + \frac{m^2+1}{r^2} + \frac{(K+\rho^2\Omega^2/A_0)}{A} + \frac{2\rho^2}{AA_0}\Omega\omega - \frac{2m}{r^2}\}. \quad (11)$$

$$U_z = -2Am(\frac{\cos\theta-1}{r^2}) + 2\frac{\rho^2}{A_0}\Omega\omega(\cos\theta-1). \quad (12)$$

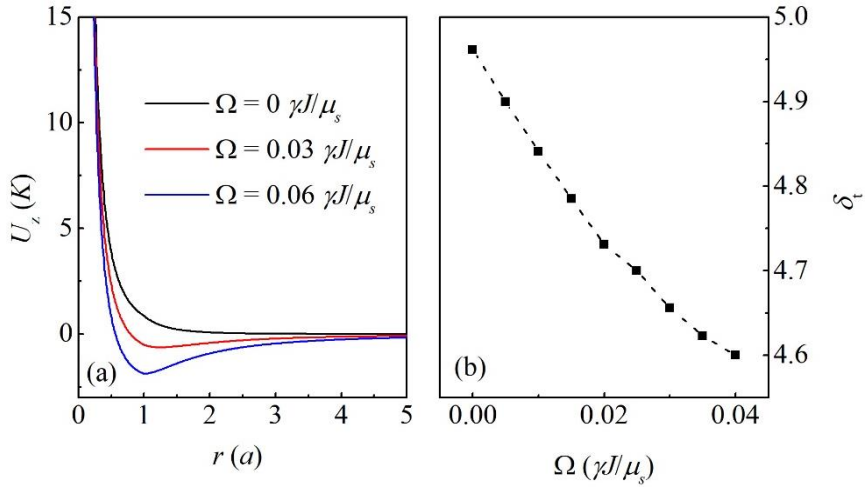

Fig. 1. (a) The calculated $U_z$ as a function of $r$ for various $\Omega$, and (b) the total scattering cross section as a function of $\Omega$.

Fig. 1(a) gives the calculated scattering potential $U_z$ as a function of $r$ for various $\Omega$ for $\omega^2/A_0 = 10K$. It is clearly shown that a local energy well in the skyrmion potential is induced by a finite $\Omega$, whose position and depth are determined by $\Omega$ and material parameters. With the increase of $\Omega$, the energy well deepens and slightly shifts towards the skyrmion center. Thus, it is expected that more magnons will be trapped in the local well when the precession is enhanced, resulting in the decrease of the scattering cross section.

For the system with skyrmion precession, the solution to the free problem $H_0 \psi_m(r) = \omega^2 \rho^2 / A_0 \psi_m$ for $\omega^2 \rho^2 / A_0 = Ak^2 + K + \Omega^2 \rho^2 / A_0 + 2\omega\Omega\rho^2/A_0$ is $\psi_m(r) = J_{m-1}(kr)$, where $J_{m-1}$ are the Bessel functions of the first kind. As $r \to \infty$, the free eigen-function $\psi^{scatter}$ reads:

$$\psi^{scatter} = (e^{i\mathbf{k}\mathbf{x}} + f(\chi)\frac{e^{ikr}}{\sqrt{r}}), \tag{13}$$

where the scattering amplitude $f(\chi)$ is defined in terms of the long-distance asymptotic behavior of the magnon wave function,

$$f(\chi) = \frac{e^{-i\pi/4}}{\sqrt{2\pi k}} \sum_{m=-\infty}^{\infty} e^{im\chi}(e^{i2\delta_{m+1}} - 1). \tag{14}$$

with the phase shift $\delta_{m+1}$. Then, one obtains the total scattering cross section

$$\delta_t = \frac{4}{k} \sum_{m=-\infty}^{+\infty} \sin^2 \delta_{m+1}. \tag{15}$$

To calculate the scattering amplitude distribution and the scattering cross section, we use the WKB approximation to estimate the phase shift $\delta_{m+1}$[30] and obtain

$$\delta_m^{WKB} = \int_{r_0}^{\infty}(\sqrt{\omega^2/A_0 - U_{eff}} - k)dr + \pi|m-1|/2 - r_0 k, \tag{16}$$

where the distance $r_0$ corresponds to the first classical turning point approaching the potential, and $U_{eff}$ reads [30]

$$U_{eff} = \frac{A(m-1)^2}{r^2} + (K + \frac{\Omega^2 \rho^2}{A_0}) + \frac{2\omega\Omega\rho^2}{A_0} + U_z. \tag{17}$$

In Fig. 1(b), we present the calculated scattering cross section $\delta_t$ as a function of $\Omega$. It is clearly shown that $\delta_t$ decreases monotonously with the increase of $\Omega$, demonstrating that more magnons are captured by the deepened potential well. Moreover, the scattering amplitude is also calculated, and the corresponding results are shown in Fig. 2(a) where gives the calculated $|f(\chi)|^2$ for various $\Omega$. For every $\Omega$, two peaks are observed, similar to that in ferromagnets [30]. Interestingly, with the increase of $\Omega$, both the peaks shift towards the large $\chi$ side, and the height of the right peak obviously decreases. Then, the scattering around the large $\chi$ region ($\sim 0.45\pi$) plays a more important role in modulating the skyrmion dynamics. As a result, the skyrmion Hall motion could be partially suppressed due to the momentum transfer between the skyrmion and magnons.

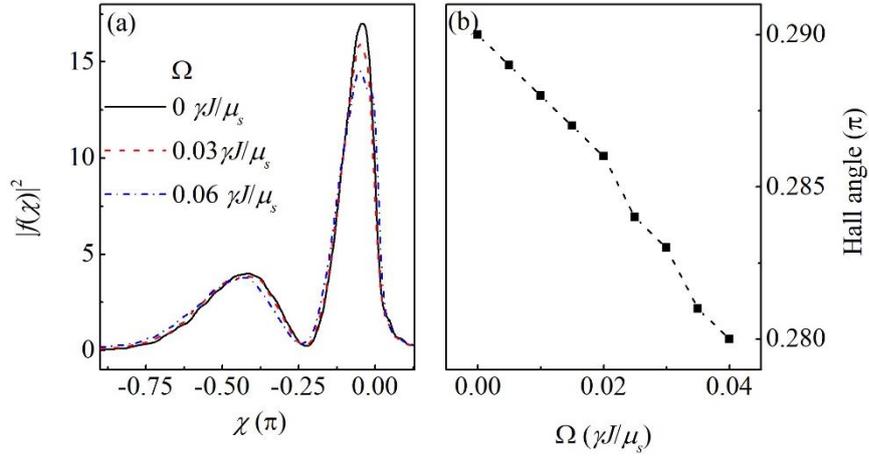

Fig. 2. (a) The calculated differential cross section as a function of $\chi$ for various $\Omega$, and (b) the skyrmion Hall angle as a function of $\Omega$.

Unlike the case in ferromagnets where the skyrmion is a massless particle and the scattering process is governed by the topology term [31], the AFM skyrmion is a Newtonian particle and its motion is mainly controlled by the inertia [32]. Thus, we consider a classic Newtonian scattering process where the skyrmion momentum is transferred from the scattered magnons, and obtain the skyrmion Hall angle depending on the scattering amplitude

$$\frac{v_y}{v_x} = \frac{\int_{-\pi}^{+\pi} -\sin(\chi)|f(\chi)|^2 \, d\chi}{\int_{-\pi}^{+\pi} (1-\cos(\chi))|f(\chi)|^2 \, d\chi}. \tag{18}$$

where $v_y$ and $v_x$ are the skyrmion speeds along the $y$-direction and the $x$-direction, respectively. The calculated Hall angle as a function of $\Omega$ is given in Fig. 2(b), which demonstrates the decrease of the angle with the increase of $\Omega$. Thus, it is theoretically predicted that the skyrmion precession affects the scattering magnitude of the injected magnons, which suppresses the skyrmion Hall motion.

C. Numerical simulations

In order to check the validity of theory, a comparison between the analytical results and numerical simulations is needed, qualitatively, at least. Thus, we also perform the numerical simulations based on the atomistic LLG equation of the synthetic AFM system, and the simulation details are explained in Appendix A. The circularly polarized magnons are generated by applying AC magnetic field with magnitude $h$.

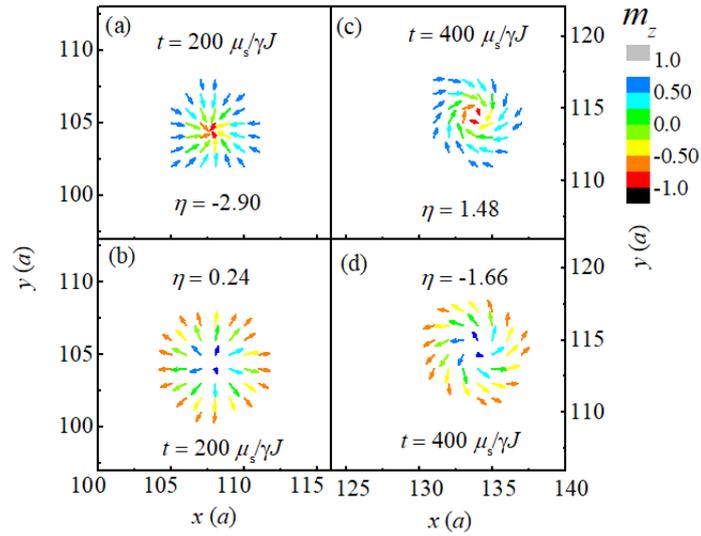

Fig. 3. The evolution of the spin configurations of the top ((a) and (c)) and bottom ((b) and (d)) layers.

Fig. 3 gives the evolution of the skyrmion configuration of the bilayer system with the damping constant $\alpha = 0.001$ which consists of a top skyrmion (a) and a bottom skyrmion (b) with opposite topology charge $Q$. The bilayer skyrmions driven by the injected magnons simultaneously move, and the skyrmion Hall motion is also observed. Moreover, the skyrmion helicity $\eta$ changes with time at a uniform speed, demonstrating a stable precession of the skyrmions. Subsequently, we investigate the dependence of the precession speed $\Omega$ on several parameters, noting that $\Omega$ is an important parameter in modulating the skyrmion dynamics.

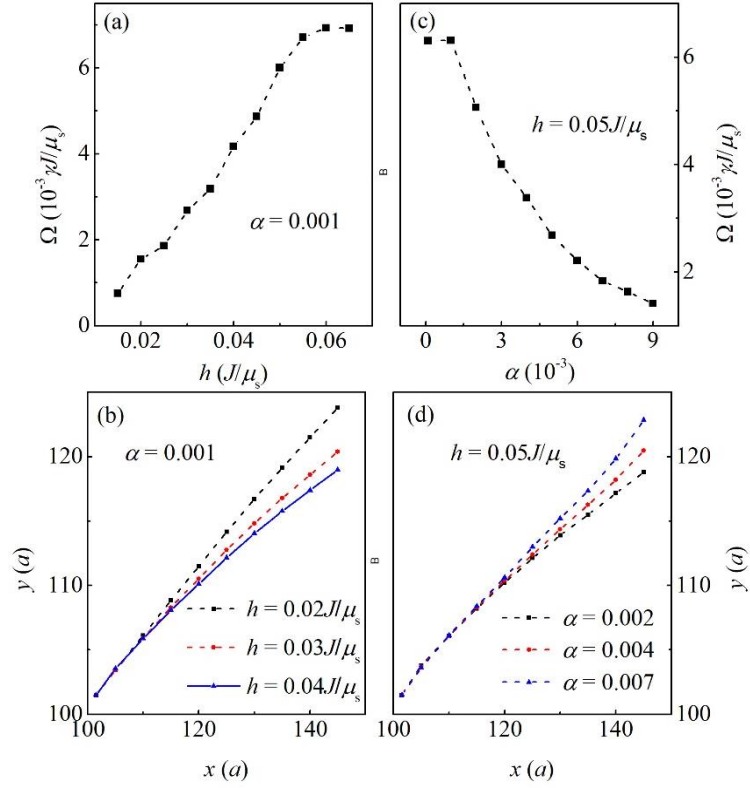

Fig. 4. The simulated (a) $\Omega$, and (b) skyrmion trajectory for various $h$ for $\alpha = 0.001$, and (c) $\Omega$, and (d) skyrmion trajectory for various $\alpha$ under $h = 0.05 J/\mu_s$.

Fig. 4(a) gives the simulated $\Omega$ as a function of $h$ for $\alpha = 0.001$. It is noted that the magnon density grows as $h$ increases, resulting in a stronger angular and linear momentum transfer between the magnons and skyrmion. Thus, the skyrmion moves faster with a larger precession speed. More interestingly, the injected magnons are strongly deflected by the skyrmion due to the changed scattering potential, resulting in the decease of the skyrmion Hall angle, as confirmed in the simulated trajectory of the skyrmion shown in Fig. 4(b). It is clearly demonstrated that the skyrmion Hall angle decreases with the increase of $h$, consistent with the theoretical prediction.

In addition, similar phenomenon is also observed in the simulated damping effect on the Hall angle. With the increasing $\alpha$, both the translation and precession speeds of the skyrmion decrease due to the fact that an enhanced damping term always lowers the skyrmion mobility, as shown in Fig. 4(c) where gives the simulated $\Omega$ as a function of $\alpha$ under $h = 0.05 J/\mu_s$. As a result, the injected magnons are weakly scattered, resulting in the increase of the skyrmion Hall angle. Thus, the Hall angle increases as $\alpha$ increases, as shown in Fig. 4(d) where gives the skyrmion trajectory for various $\alpha$.

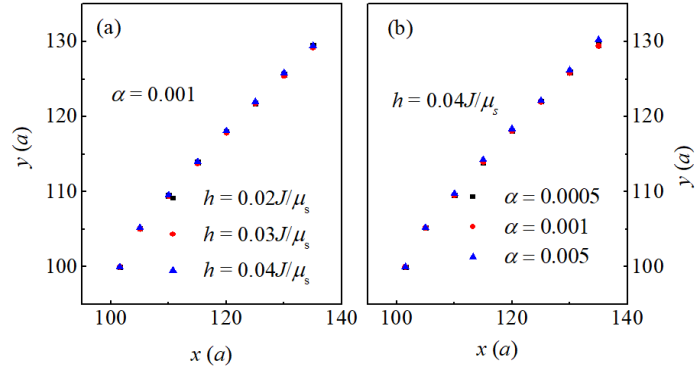

Fig. 5. Skyrmion trajectory of the system without skyrmion precession (a) for various $h$ for $\alpha = 0.001$, and (b) for various $\alpha$ under $h = 0.04 J/\mu_s$.

So far, the effect of the skyrmion precession on the dynamics has been analytically and numerically revealed in the frustrated system. However, one may note that the skyrmion precession does not exist in magnets with inversion broken symmetry. Thus, the fictitious electromagnetic field from the skyrmion is unaffected by the AC magnetic field and damping constant, resulting in the independence of the skyrmion Hall angle on these parameters. This property is confirmed in our simulations, as shown in Figs. 5(a) and 5(b) where present the skyrmion trajectories of the system with an additional DMI for various $h$ and various $\alpha$, respectively. When the DMI is introduced (the simulation details are provided in Appendix C), the skyrmion precession is completely suppressed, and the skyrmion Hall angle is not affected by $h$ and $\alpha$.

*D. Brief discussion*

Different from the earlier work [17], the $(\nabla^2 n)^2$ term in the continuum Hamiltonian is ignored in the theoretical calculations, considering the fact that we mainly focus on the influence of the precession on the skyrmion dynamics. Moreover, the energy contribution from the $(\nabla^2 n)^2$ term is much less than ($\sim 1/6$) that from the $(\nabla n)^2$ in the studied system, while the consideration of the $(\nabla^2 n)^2$ term makes the scattering equation too complex to be solved clearly.

Importantly, it is worth noting that the ignorance of the term hardly affects our main conclusion, as confirmed in the numerical simulations.

Thus, it is revealed that the skyrmion precession suppresses the Hall motion, and the skyrmion Hall angle can be effectively modulated by the injected magnon intensity and the damping constant in frustrated magnets, providing useful information for future spintronic and magnonic applications. In addition, the main conclusion of the AFM bilayer system can naturally extend to frustrated ferromagnets and ferrimagnets where the berry phase are replaced by $\cos\theta\partial_t\varphi$ and $\delta_s\cos\theta\partial_t\varphi + \rho_0\dot{n}^2/2$ respectively. Furthermore, for two-sublattice systems where the linearly polarized magnons can be generated, the skyrmion precession direction and the eigen-function should be updated, while the essential physics is similar.

## V. Conclusion

In conclusion, we have studied analytically and numerically the dynamics of the frustrated skyrmion in synthetic antiferromagnets driven by the circularly polarized magnons. It is revealed that the scattering cross section of the magnons depends on the skyrmion precession speed, which affects the skyrmion Hall motion. Interestingly, the Hall angle decreases with the increase of the precession speed which can be modulated by the magnon intensity and the damping constant, providing an effective way in suppressing the Hall motion. Thus, the present work unveils new skyrmion-magnon scattering mechanisms in frustrated magnets, benefiting future spintronic and magnonic applications.

## Acknowledgment

We sincerely appreciate the insightful discussions with Huaiyang Yuan. The work is supported by the Natural Science Foundation of China (Grants No. 51971096 and No. 51721001), and the Natural Science Foundation of Guangdong Province (Grant No. 2019A1515011028), and the Science and Technology Planning Project of Guangzhou in China (Grant No. 201904010019).

**Appendix A: The derivation of the Lagrangian density**

We consider two FM layers with competing Heisenberg exchange interactions based on the $J_1$–$J_2$–$J_3$ model on a 200 × 200 square lattice [23].

$$H = H_{top} + H_{bottom} + H_{inter}$$
$$H_{top,bottom} = -J_1 \sum_{\langle i,j \rangle} \mathbf{m}_i^{1,2} \cdot \mathbf{m}_j^{1,2} - J_2 \sum_{\langle\langle i,j \rangle\rangle} \mathbf{m}_i^{1,2} \cdot \mathbf{m}_j^{1,2} - J_3 \sum_{\langle\langle\langle i,j \rangle\rangle\rangle} \mathbf{m}_i^{1,2} \cdot \mathbf{m}_j^{1,2} - K_0 \sum_i (\mathbf{m}_i^{1,2} \cdot \mathbf{z})^2$$
$$H_{inter} = -J_{inter} \sum_i \mathbf{m}_i^1 \cdot \mathbf{m}_i^2 \qquad (A1)$$

where $J_1 = J$ is the exchange interaction between the nearest neighbor spins, $J_2 = -0.1J$ and $J_3 = -0.15J$ are the interactions between the next-nearest neighbors and next-next-nearest neighbors respectively, and $K_0$ is the anisotropy constant, and $J_{inter}$ is the AFM interface coupling between the nearest neighbors [45].

Considering the normalized staggered Néel vector $\mathbf{n} = (\mathbf{m}^1 - \mathbf{m}^2)/2S$ and magnetization $\mathbf{m} = (\mathbf{m}^1 + \mathbf{m}^2)/2S$, we can rewrite the Hamiltonian

$$H = -2J_1 \sum_{\langle i,j \rangle} \mathbf{n}_i \cdot \mathbf{n}_j - 2J_2 \sum_{\langle\langle i,j \rangle\rangle} \mathbf{n}_i \cdot \mathbf{n}_j - 2J_3 \sum_{\langle\langle\langle i,j \rangle\rangle\rangle} \mathbf{n}_i \cdot \mathbf{n}_j - 2K_0 \sum_i (\mathbf{n}_i \cdot \mathbf{z})^2 + J_{inter} \sum_i \mathbf{m}_i^2, \qquad (A2)$$

and Taylor expand $\mathbf{n}_j = \mathbf{n}(ja)$ around $\mathbf{n}(ia)$. Then, the exchange energy expression treated within the lowest relevant order reads [46]

$$H_{ex} = \sum_i \Delta \{ \sum_{j>i} [-\frac{1}{2}(j-i)^2 J_{j-i}(\nabla \mathbf{n})^2] \}. \qquad (A3)$$

Using the continuity approximation, one can rewrite the Eq. (A2)

$$H = \frac{A}{2}(\nabla \mathbf{n})^2 + \frac{K}{2} n_z^2 + \frac{A_0}{2} m^2, \qquad (A4)$$

where $A = (J_1 - 2J_2 - 4J_3)$, $A_0 = 2J_{inter}$, and $K = 4K_0$.

**Appendix B: Numerical simulations of the atomistic spin model**

The dynamics of the skyrmion driven by the magnons is investigated by solving the LLG equation,

$$\frac{\partial \mathbf{S}_i}{\partial t} = -\gamma \mathbf{S}_i \times \mathbf{H}_i + \alpha \mathbf{S}_i \times \frac{\partial \mathbf{S}_i}{\partial t}, \qquad (B1)$$

where $\alpha$ is the damping constant, $\mathbf{H}_i = -\mu_s^{-1}\partial H/\partial \mathbf{S}_i$ is the effective field. Without loss of generality, $K_0 = 0.02\,J$, and $J_{\text{inter}} = -0.7J$ are selected. Generally, we use the fourth-order Runge-Kutta method to solve the LLG equation on a $200 \times 200$ square lattice.

The magnons are excited in the region $x = [60, 64]$ by a homogeneous and dimensionless magnetic field source. Here, we generated the circularly polarized magnons, by applying an AC magnetic field $h_{\text{RH}} = h\,[\cos(\omega_0 t)\,\mathbf{x} + \sin(\omega_0 t)\,\mathbf{y}]$ with $\omega_0 = 0.6\gamma J/\mu_s$. To exclude the reflection of magnons at boundary, we impose absorbing boundary conditions. In this work, the position of the skyrmion $X_i$ is estimated by

$$X_i = \frac{\int [i\mathbf{n}\cdot(\partial_x\mathbf{n}\times\partial_y\mathbf{n})]dxdy}{\int [\mathbf{n}\cdot(\partial_x\mathbf{n}\times\partial_y\mathbf{n})]dxdy}, i = x, y. \tag{B2}$$

Then, the velocity is numerically calculated by $v = dX_i/dt$ with time $t$.

**Appendix C: The consideration of the DMI**

In order to suppress the skyrmion precession, an additional DMI between the in-plane nearest neighbors is considered,

$$H_{\text{DMI}} = -D_0 \sum_i (\mathbf{m}_i^{1,2}\times\mathbf{m}_{i+x}^{1,2}\cdot\mathbf{e}_y - \mathbf{m}_i^{1,2}\times\mathbf{m}_{i+y}^{1,2}\cdot\mathbf{e}_x), \tag{C1}$$

with the coupling $D_0 = 0.17J$. Moreover, $J_2 = J_3 = 0$, $J_{\text{inter}} = -0.7J$, $K_0 = 0.05J$ is considered for simplicity.